\documentclass{iaus}
\usepackage{graphicx}

\title[SPHOTOM package] 
{SPHOTOM -- Package for an Automatic Multicolour Photometry}

\author[\v{S}. Parimucha, M. Va\v{n}ko \and P. Miklo\v{s}]   
{\v{S}. Parimucha$^1$, M. Va\v{n}ko$^2$ \and P. Miklo\v{s}$^1$}

\affiliation{$^1$Institute of Physics, University of P.J. \v{S}af\'arik, Ko\v{s}ice, Slovakia \\ email: {\tt stefan.parimucha@upjs.sk} \\[\affilskip]
$^2$Astronomical Institute of the Slovak Academy of Sciences, Tatransk\'a Lomnica, Slovakia  \\email: {\tt vanko@ta3.sk}}

\pubyear{2011}
\volume{xxx}  
\pagerange{119--126}
\jname{Title of your IAU Symposium}
\editors{A.C. Editor, B.D. Editor \& C.E. Editor, eds.}
\begin{document}

\maketitle

\begin{abstract}
We present basic informations about package SPHOTOM for an automatic multicolour photometry. This package is in development for a creation of photometric pipe-line, which we plan to use in near future with our new instruments. It could operate in two independent modes, (i) GUI mode, in which user can select images and control functions of package through interface and (ii) command line mode, in which all processes are controlled using a main parameter file. SPHOTOM is developed as an universal package for Linux based systems with easy implementation for different observatories. Photometric part of package is based on Sextrator code, what allow us to detect all objects on the images and perform their photometry with different apertures. We can also perform astrometric solution for all images for a correct cross-identification of the stars on the images. The result is a catalogue of all objects with their instrumental photometric measurements which are consequently used for a differential magnitudes calculations with one ore more comparison stars, transformations to international system and colour indices determinations.

\end{abstract}


\noindent The function of the SPHOTOM can be described in the following steps:

\subsection*{Sorting}
It is the first step in command line mode, which creates different lists of images based on informations in FITS header of 
images as well as names of files. It uses robust sorting scheme and it is written in Python using PyFits module. User can define
types of lists, which will be used in next steps. Sorting could be executed also from GUI mode. For a correct functionality it
is necessary to have a consistent FITS headers and/or image names, which are observatory dependent. 

\subsection*{Master images}
Create master images using an average or a median of input files. No other corrections are performed in this step. This images are stored in archive for a future use. User can define a number of images entering into this procedure.

\subsection*{Photometric reduction}

During photometric reduction of raw images we use created or archive master images. We can use, bias, dark frame, flat-field and dark for a flat corrections. Procedure automatically control image dimensions, temperatures and used colours. The results are the lists of images based on their colours.

\subsection*{Photometry}

All images in lists after photometric reduction are used for a photometry using Sextrator (Bertin \& Arnouts, 1996) code. We can control all photometry options using parameters files of Sextrator (for more details see Sextrator manual). This package is very effective on non or semi-crowded fields and allow us to detect all objects above defined background level on the images and perform their photometry with different user defined apertures. It can also remove bad or corrupted detections (stars on the edges of images, saturated stars, cosmic ray hits). User can define different types of informations in output file.

\subsection*{Identification}

If FITS header of images have a WCS (World Coordinate Systems) information (Calabretta \& Greisen, 2002), we perform cross-identification with external catalogs (USNO-A2.0 or Tycho) using up to 20 brightest stars on image and than astrometric solution of all detected objects is calculated. If we have no WCS information we calculate astrometric solution with known approximate coordinates of image center and cross-Identification with external catalogs. For each image it is created a file with detected objects with their coordinates (celestial and image) and their instrumental magnitudes in apertures from photometry. 

\subsection*{Output catalogue}

After identification of the objects  we have 2 possibilities: \newline
(I) in GUI user can select up to 9 stars and generate their multicolour light curves with differential magnitudes. No other corrections are performed. User can manually use procedures in final corrections step. \newline
(II) in command line mode we generate differential magnitudes between all pairs of stars, create light curves and  store them in temporary database for an easier manipulation.

\subsection*{Final corrections}

In final step we can perform several corrections: \newline
1) Heliocentric correction of time. \newline
2) Determination of extinction coefficients and reduction of systematic effects using SARS algorithm (Ofir et al., 2010). \newline 
3) Transformation to international system with known transformation coefficients. \newline
4) Calculation of average comparison star from user selected objects. \newline
Finally we create result differential light curves.

This contribution was supported by VEGA project 2/0094/11.

\end{document}